\documentclass[prl,twocolumn,superscriptaddress]{revtex4-2}
\usepackage{natbib}
\usepackage{amsmath}
\usepackage{amsfonts}
\usepackage{graphicx}
\usepackage{dcolumn}
\usepackage{tabularx}
\usepackage{hyperref}
\usepackage{keyval}
\usepackage{multirow}
\usepackage{xcolor}
\usepackage{CJKutf8}

\begin{document}
\begin{CJK*}{UTF8}{bsmi}
\title{Optical response of the bulk stabilized mosaic phase in Se doped TaS$_{2-x}$Se$_{x}$.}

\author{Xuanbo Feng (馮翾博)}
\affiliation{van der Waals - Zeeman Institute, University of Amsterdam, Sciencepark 904, 1098 XH Amsterdam, the Netherlands}
\affiliation{QuSoft, Science Park 123, 1098 XG Amsterdam, The Netherlands}

\author{Liam Farrar}
\affiliation{Centre for Nanoscience and Nanotechnology, Department of Physics, University of Bath, Bath BA2 7AY, United Kingdom} 

\author{Charles J. Sayers}
\affiliation{Centre for Nanoscience and Nanotechnology, Department of Physics, University of Bath, Bath BA2 7AY, United Kingdom} 
\affiliation{Dipartimento di Fisica, Politecnico di Milano, 20133 Milano, Italy}

\author{Simon J. Bending}
\affiliation{Centre for Nanoscience and Nanotechnology, Department of Physics, University of Bath, Bath BA2 7AY, United Kingdom} 

\author{Enrico Da Como}
\affiliation{Centre for Nanoscience and Nanotechnology, Department of Physics, University of Bath, Bath BA2 7AY, United Kingdom}

\author{Erik van Heumen}\email{e.vanheumen@uva.nl}
\affiliation{van der Waals - Zeeman Institute, University of Amsterdam, Sciencepark 904, 1098 XH Amsterdam, the Netherlands}
\affiliation{QuSoft, Science Park 123, 1098 XG Amsterdam, The Netherlands}

\date{\today}

\begin{abstract}
 The layered van der Waals material, TaS$_{2}$ features a meta-stable mosaic phase on the verge of a nearly commensurate to commensurate charge density wave transition. This meta-stable or 'hidden' phase can be reached by laser pumping the low temperature, commensurate charge density wave phase. Here we report the stabilization of a bulk, equilibrium mosaic phase in 1T-TaS$_{1.2}$Se$_{0.8}$ single crystals observed with transport and optical spectroscopy experiments. We identify a bulk pseudogap in the mosaic phase of approximately 200 meV at the lowest temperatures, while the CCDW phase can be obtained by heating and instead has a full optical gap of about 100 meV. Surprisingly, a spectral weight analysis shows that Se doping gives rise to an increased charge density despite the fact that this is formally an isovalent substitution. This finding is consistent with the recent observation that the mosaic phase is stabilized as equilibrium phase through the appearance of charged defects.  
\end{abstract}
\maketitle
\end{CJK*}

Transition metal dichalcogenides (TMDCs) provide a rich playground of exotic electronic ordering phenomena such as superconducting or charge density wave phases. 1T-TaS$_{2}$ stands out among them due to the observation of a correlated insulating phase, the nature of which is still under debate \cite{Ang2013, Qiao2017, Ritschel2018, Lee2019, Butler2020, Wang2020, Petkov2022}. Disentangling the interplay between structural and electronic order is difficult, but the consensus now appears to be that the equilibrium groundstate is a commensurate charge density wave that originates in dimerization of layer stacking along the c-axis \cite{Ritschel2018, Lee2019}. However, at elevated temperatures \cite{Wang2020} or at specific surface terminations \cite{Butler2020} signatures of Mott physics reemerge pointing to the interplay of several instabilities. Against this backdrop, a series of experiments employing light or current pulses demonstrated the existence of a `hidden' or meta-stable state \cite{Stojchevska2014, Han2015, Gerasimenko2019, Ravnik2021, Gao2021, Venturini2022}. The induced metallicity appears to be tied to the emergence of domain walls in the CCDW phase \cite{Ma2016,Cho2016,Gerasimenko_npj_2019} and changes in the interlayer stacking \cite{Stahl2020}. The meta-stable state has subsequently been dubbed the `mosaic' phase. The tunability between distinct electronic phases with external control has led to suggestions for applications \cite{Yoshida2015, Mihailovic2021} and consequently stabilising the mosaic phase under equilibrium conditions would be very useful. Recently, an equilibrium version of the mosaic phase was reported to exist as a surface effect on 1T-TaS$_{2}$ single crystals \cite{Salzmann2022}. Similar equilibrium states appear to exist as surface structure in doped 1T-TaS$_{2-x}$B$_{x}$ crystals where B is Ti \cite{Zhang2022} or Se \cite{Endo2000, Qiao2017, Fujii2018}. 

Here we report the existence of a meta-stable, metallic phase in bulk single crystals of Se doped 1T-TaS$_{2-x}$Se$_{x}$ (x\,=\,0.8) that can be reached under thermal equilibrium conditions. This state is observed in a large temperature window below 130\,K and is distinctly different from the insulating CCDW phase. The latter can be realized after heating above 130\,K and it remains the equilibrium state if the crystal is subsequently cooled down again. From a detailed analysis of the optical spectra, we conclude that there are two distinct mechanisms that give rise to the mosaic and CCDW phase. The former is characterized by the formation of a pseudogap, while the latter is characterized by a full optical gap.   

Single crystals of 1T-TaS$_{2-x}$Se$_{x}$ with nominal compositions x\,=\,0.8 and x\,=\,1.0 are are grown using chemical vapor transport. Typical crystal sizes obtained in this way are approximately 2\,$\times$\,3\,mm with a somewhat irregular shape. Further details of the growth process are provided in the supplementary material. Transport experiments were carried out using a 4-point method using a Physical Property Measurement System (PPMS) and data was recorded between 3 and 310 K. Infrared reflectivity spectra are collected over the energy range from 6 meV to 4 eV and between 16 K and 400 K for two samples with Se contents x\,=\,0.8 and x\,=\,1.0. In order to study the hysteresis in this system, a temperature loop is designed to start cooling from 400 K down to 16 K. At base temperature, data is collected for 20 minutes and then the sample is heated to 400 K. Both cooling and warming experiments are carried out with a constant temperature change of 1.7 K/min. Each cycle is repeated several times to verify the reproducibility and reduce the noise in the data. The experiment is repeated right after the \textit{in situ} evaporation of silver or gold on the samples to obtain reference spectra. These allow us to determine the absolute reflectivity of our samples. To obtain the optical conductivity, we use the variational dielectric function method proposed in Ref. \cite{Kuzmenko2005}.

\begin{figure}[thb]
    \includegraphics[width=0.99\columnwidth]{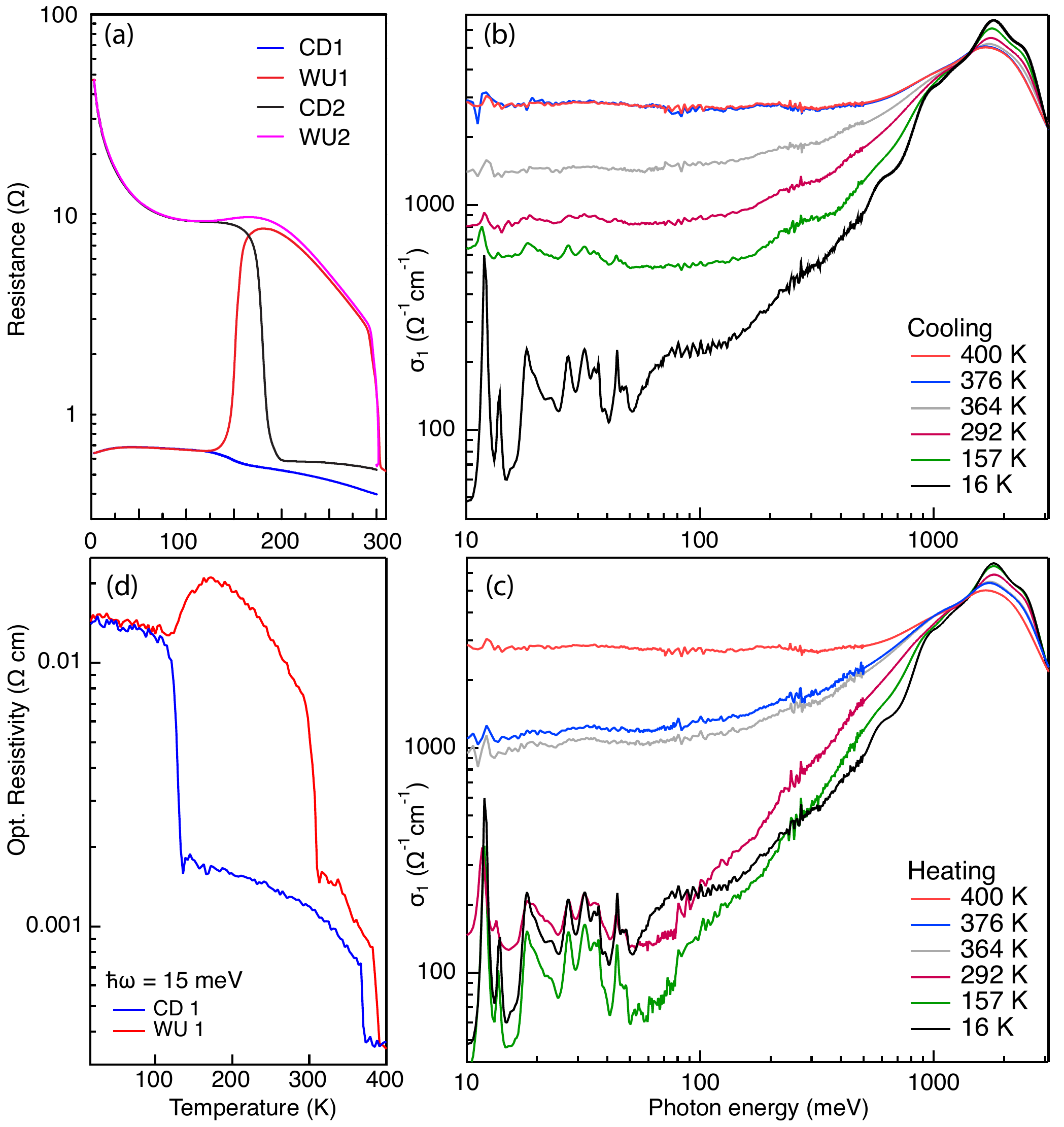}
    \caption{Observation of a bulk mosaic phase in 1T-TaS$_{1.2}$Se$_{0.8}$. (a): Resistance cycles between 300 K and 1 K. As the sample is cooled from high temperature a first time (CD 1, blue curve), a weak semi-conducting like behaviour is observed. Upon heating, a transition to a more insulating phase is observed (WU 1, red curve). As long as temperature is kept below 380\,K, the temperature dependent resistance now resembles that of pure 1T-Ta$S_{2}$ (CD 2/WU2, black and magenta curves). (b,c): Optical conductivity of TaS$_{1.2}$Se$_{0.8}$ measured during cooling (b) and heating (c). As temperature decreases a series of IR active phonon modes become visible. Clear spectral differences appear in the optical conductivity data during heating compared to the cooling data.This is further exemplified in panel (d), which shows the optical resistivity ($[\sigma_{1}(15\, meV,T)]^{-1}$). The optical resistivity closely tracks the resistance of panel (a) and demonstrates that the sample can be rest when heated above 380\,K.}
    \label{fig:s1}
\end{figure}
A first investigation of the bulk electronic properties is carried out with transport experiments, Fig. \ref{fig:s1}a. The sample is briefly heated to 400\,K and subsequently mounted in the PPMS for resistance measurements. These single crystals are relatively thin and have a somewhat irregular shape. This prevented us from accurately determining the resistivity and we therefore only show the measured resistance. As the crystal is cooled a first time, we observe a weak semi-conducting behaviour of the resistance (CD1, blue curve). Around 130 K, a change in slope can be observed that signals a phase transition. As temperature is further reduced, the resistance eventually starts to decrease at the lowest measured temperatures. As the crystal is heated, we again observe a transition around 130 K (WU1, red curve). Surprisingly, the resistance increases by an order of magnitude before passing through a maximum. A new transition to a low resistance state appears around 305\,K. Additional cooling and heating cycles display markedly different behaviour to the first cycle. We now observe a phase transition at 180\,K to an insulating phase as evidenced by the large upturn at low temperature (CD2, black curve). The second heating cycle (WU2, magenta curve) closely tracks the cooling cycle CD2, but displays a large hysteresis in the phase transition (305\, K heating compared to 180\,K cooling). The difference between the insulating (WU2) and metallic (WU1) behaviour closely resembles the changes in resistivity that take place when light pulses induce the mosaic phase in the CCDW phase of 1T-TaS$_{2}$ \cite{Stojchevska2014}, which is a first indication that the mosaic phase is stabilized under equilibrium conditions in our crystals.

To further explore this, we turn to the real part of the optical conductivity, $\sigma_1(\omega,T)$ in Fig.\,\ref{fig:s1}. Data is presented separately for cooling and heating to highlight the distinct behaviour of the optical response. The optical conductivity of the incommensurate charge density wave phase (ICCDW; T$\,\sim\,$400 K) is characterized by a nearly frequency independent free charge response. Compared to the optical conductivity of pristine 1T-TaS$_2$, reported in Ref.\,\cite{Gasparov2002} and Ref.\,\cite{velebit2015}, the extrapolated DC conductivity of Se doped crystals is similar. The interband conductivity above 1 eV is significantly larger in our crystals, although the three interband transitions at 1 eV, 1.66 eV and 2.2 eV are also observed in previous measurements. We further note that the interband response show a significant temperature dependence.

At the transition to the nearly commensurate state (NCCDW; T$_{NCCDW}\,\approx\,$370 K), we observe a sudden depletion of spectral weight below 1 eV (visible as a step in the temperature dependent optical resistivity in Fig.\,\ref{fig:s1}(d)). This depletion evolves similarly for a crystal with x\,=\,1.0 (see Supplementary material). Different from the x\,=\,1.0 crystal, the $x\,=\,0.8$ crystal undergoes a second transition at 130 K where another sudden removal of spectral weight takes place. For 1T-TaS$_{2}$, this transition has been identified as the formation of the commensurate CDW phase (CCDW). As we will discus below, the $x\,=\,0.8$ crystal first enters an intermediate meta-stable phase.

As the screening from free charge carriers is reduced, a series of phonon modes becomes visible. A group theoretical analysis predicts 40 infrared active phonon modes in the CCDW phase of 1T-TaS$_{2}$ \cite{Gasparov2002}. This number can be expected to become even larger for Se doped crystals, since the substitution of some of the S atoms with Se atoms will lead to mode splitting and frequency shifts. This is indeed what we observe: there are two additional phonon modes at 18 meV and 19.6 meV that are not present in the earlier infrared data\cite{Gasparov2002, velebit2015}. As was pointed out in Ref. \cite{velebit2015}, the frequency splitting between the phonon modes can be very small. Our experimental resolution of 0.25 meV is certainly not sufficient to observe all possible infrared active modes. Apart from the extra phonon modes, we observe a similar number of modes compared to earlier work although they are all shifted in frequency. Most modes appear broader and we attribute this to unresolved mode splitting due to Se substitution.

We now return to the second transition of the $x\,=\,0.8$ crystal at 130 K. In 1T-TaS$_{2}$ a transition from the NCCDW to CCDW phase takes place at 180 K. Angle resolved photoemission spectroscopy (ARPES) and scanning tunneling microscopy (STM) experiments have shown that this transition is accompanied by the opening of a large gap at the Fermi level and the formation of bands below and above the Fermi level \cite{Zwick1998, Clerc2006, Wang2020, Butler2020}. Regardless the nature of this transition, it has been shown that the resistivity below this transition becomes insulating with an exponential enhancement of the resistivity, $\rho(T)\propto exp(-A/k_{B}T)$ \cite{DiSalvo1977}. 

Our optical conductivity data indeed shows the formation of a gap, but still has a significant background conductivity at 16 K. We attribute this to a residual metallicity that would be expected for the mosaic phase and is consistent with the low temperature resistance data (CD1). Similar to the resistance data (WU1), we observe a further depletion of spectral weight when the crystal is heated above 130 K. The 157\,K data in Fig. \ref{fig:s1}c shows a clear suppression of low energy spectral weight compared to the 16\,K data. We can directly identify different phases observed in transport experiments with optical spectra by comparing resistance with optical resistivity data, Fig. \ref{fig:s1}d. The optical resistivity (inverse of the optical conductivity) shows quantitative differences with the resistance as might be expected given that the former is measured at finite energy and in the phonon range. Nevertheless, at $\hbar\omega\,=\,15$\,meV it compares qualitatively very well with the measured resistance and we can map the optically observed transitions one-to-one to those in the resistance. The residual conductivity observed in the optical data at 16\,K is a second indication that a mosaic phase is stabilized in 1T-TaS$_{1.2}$Se$_{0.8}$. Since optical experiments probe the volume of the crystal, this is a clear demonstration that we are observing a \textit{bulk} mosaic phase and not just a surface effect. 

The temperature dependence of the optical conductivity provides another clue that this phase is indeed meta-stable. During our experiment, we stabilize the temperature after cooling to 16 K for 20 minutes. We observe a small but significant change in the optical response that is most visible in Fig.\,\ref{fig:s1}(d) (a small difference between the red and blue curve; a more prominent difference is observed at higher photon energy, see supplementary Fig. \ref{fig:sups1}d). 

Based on the above observations, we thus identify the 16 K optical conductivity as representative for the electronic spectrum of the mosaic phase. The lowest temperature optical conductivity spectrum of the CCDW phase is in our experiment only obtained above 130 K after heating the sample (Fig.\,\ref{fig:s1}(b); green curve). We can exclude that this transition is to one of the other CDW phases that have been observed in 1T-TaS$_{2}$ \cite{Tanaka1984,Suzuki1988,Burk1992}: as temperature increases further, three more transitions are visible in our data. These resemble transitions to the T-phase (305 K), NCCDW phase (340 K) and ICCDW phase (385 K). 

The question that now emerges is whether there is a difference in the nature of the transitions between NCCDW and mosaic phase and between the mosaic and CCDW phase. These questions are often approached by making use of spectral weight analysis. However, in many TMDC's, it has been observed that the temperature dependence of the optical spectra is significant even in the visible and UV parts of the optical spectrum \cite{Whitcher2021}. It has been speculated that this is a consequence of changes in interlayer coupling that emerge when lattice expansion or contraction takes place \cite{Tanaka1984, Wang2019}. The temperature dependence of 1T-TaS$_{2-x}$Se$_{x}$ appears to follow this trend. Our optical conductivity data shows significant temperature dependence over the entire spectral range, up to 2 eV. These significant changes in interband spectra perhaps find their origin in a reduction of the c-axis lattice constant as temperature is reduced. A significant portion of this spectral weight enhancement thus may be unrelated to the various charge density wave transitions and we have no way to disentangle this 'trivial' spectral weight change from redistributions due to the opening of new charge density wave related gaps.

\begin{figure*}[thb]
    \includegraphics[width = 2.0\columnwidth]{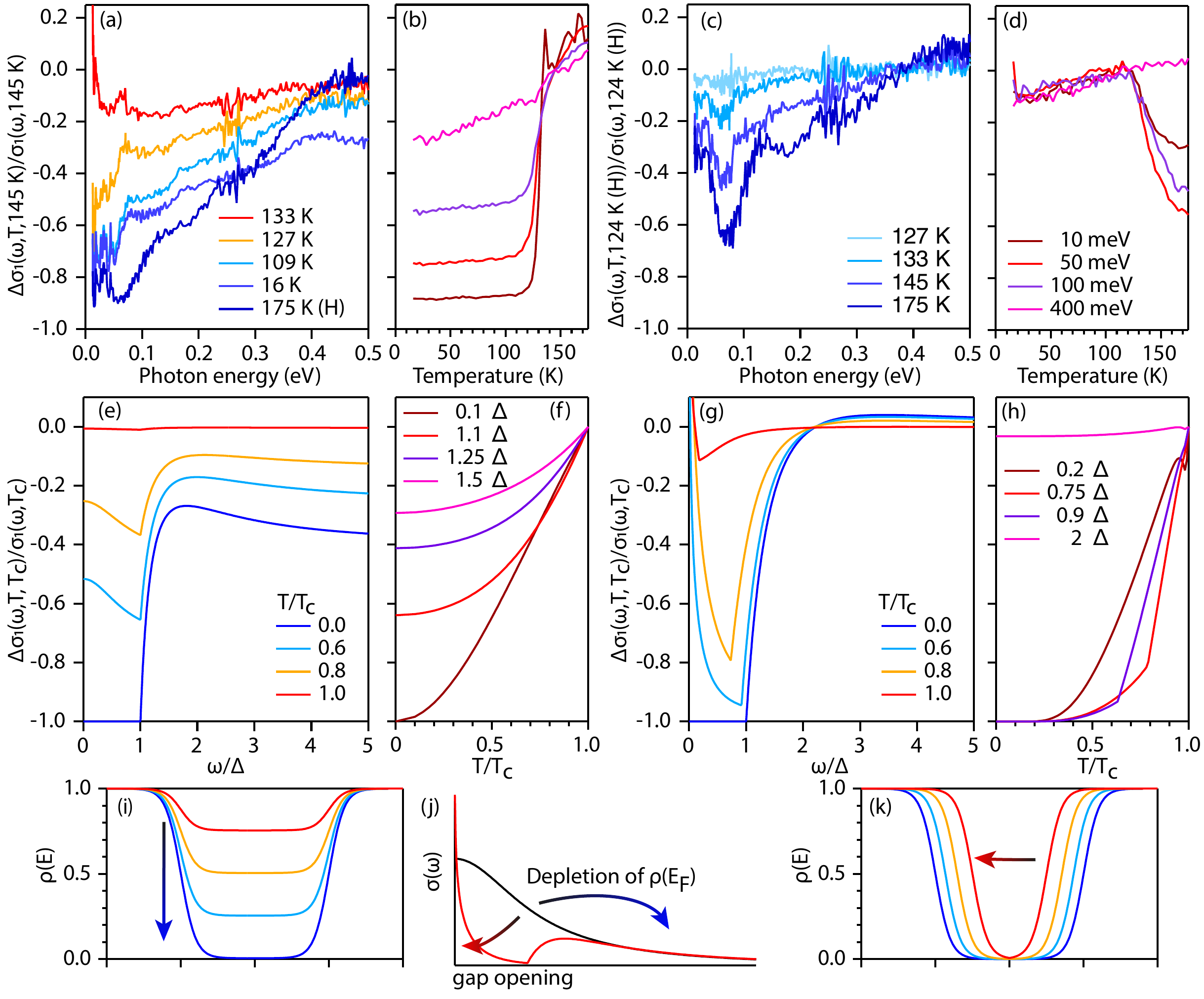}
    \caption{(a): The optical conductivity difference $\Delta\sigma_1(\omega, T,T_{0}=\,145\,K)$ during cooling at selected temperatures. All temperatures are measured on cooling except the curve labelled 175 K (H). (b): As temperature decreases, a depletion in the optical response forms with the largest change taking place at the lowest energy (legend same as panel (d)). (c): The optical conductivity difference $\Delta\sigma_1(\omega, T,T_{0}=\,124\,K)$ during heating at selected temperatures. (d): The largest change in $\Delta\sigma_1(\omega, T,T_{0})$ now takes place at finite energy (50 meV). (e,f): the temperature dependence during cooling can be qualitatively understood from the removal of free charge spectral weight on top of a gapped background conductivity. (g,h): the changes observed during heating are better described by the opening of a gap in the optical conductivity. (i): when the density of states around the Fermi energy is depleted, a transfer of spectral weight (panel j) to high energy takes place. (k): when a gap opens, transfer to low energy can also occur. See main text for more detailed discussion.}
    \label{fig:ds}
\end{figure*}

We therefore turn to another quantity to highlight differences in the formation of the mosaic and CCDW phases: the normalised difference of the optical conductivity. In Fig.\,\ref{fig:ds}, we calculate the difference between the optical conductivity and a reference temperature, normalised to the conductivity at the same reference temperature, $\Delta\sigma_1(\omega,T,T_{0})/\sigma_1(\omega,T_{0})$. Having identified the transition temperature of the NCCDW to the mosaic phase around T$_{0}\,\approx\,$145 K from Fig.\,\ref{fig:s1}(c), we plot in Fig.\,\ref{fig:ds}(a) $\Delta\sigma_1(\omega,T,145\,K)/\sigma_1(\omega,145\,K)$. With decreasing temperature the spectral weight at low energy is depleted, with the largest changes happening at the lowest photon energy. As the photon energy increases, the changes become gradually smaller, as can be more clearly seen from Fig. \ref{fig:ds}b. 

The transition from the mosaic to the CCDW state during heating follows a different behaviour. To highlight this, we take T$_{0}\,\approx$ 124 K as reference. Fig.\,\ref{fig:ds}c shows that a pronounced minimum develops around 0.1 eV and possibly a second minimum around 0.2 eV. Fig.\,\ref{fig:ds}(d) shows that the largest change in the optical response now takes place at \textit{finite} photon energy between 50-100 meV. This difference in temperature dependence between cooling and heating points to different gap formation mechanisms as we will discuss next.  

The two likely scenarios along which a gap opens at the Fermi level are a depletion (Fig. \ref{fig:ds}i) or gradual opening of a gap (Fig. \ref{fig:ds}k) of the density of states around the Fermi level. The later case is often associated with spontaneous symmetry breaking phases that are accompanied by the formation of a temperature dependent gap and associated Goldstone modes (sliding modes in this case). The temperature dependent optical response of such a BCS type phase transition was numerically evaluated by Zimmerman et al. in Ref. \cite{Zimmermann1991} and has been implemented in the software package RefFit \cite{kuzmenko:reffit}. We model the temperature dependent response of $\Delta\sigma(\omega,T,T_{c})$ using this numerical code and plot the result in Fig. \ref{fig:ds}g. The temperature dependence at selected energies is shown in Fig. \ref{fig:ds}h. A comparison between these panels and corresponding experimental panels (Fig. \ref{fig:ds}c and d respectively), shows qualitatively similar behavior. 

The impact of a depletion in the density of states on the optical response is harder to model: it requires a concrete theoretical backing of the phenomenon or one has to resort to approximate estimates making use of the joint density of states (JDOS). The advantage of the JDOS is that an approximate estimation of the optical response can be used for both gap opening and gap closing scenarios. However, since we have an exact method available for the BCS case, we use a similar approach to model the pseudogap formation. This is achieved by taking a sum of a Drude term and the T=0 BCS optical conductivity:
\begin{equation}\label{PGmod}
\sigma(\omega,T)=(T/T_{c})^2\sigma_{Dr.}(\omega,T)+(1-T/T_{c})^2\sigma_{BCS}(\omega,0)
\end{equation} 
We have verified that this gives the same qualitative result as the JDOS approximation. Eq. \ref{PGmod} allows us to introduce a small temperature dependence in the Drude response, in particular in the Drude width $\Gamma(T)$. The result of these simulations is shown in Fig. \ref{fig:ds}e,f. The results qualitativel reproduce the experimental results in Fig. \ref{fig:ds}a,b. The depletion of $\Delta\sigma_1(\omega,T,145\,K)/\sigma_1(\omega,145\,K)$ above 0.1 eV is reproduced in panel \ref{fig:ds}e for energies $\omega/\Delta\,>\,1$. This requires us to assume a temperature dependent Drude component for which the $\Gamma(T)$ decreases. The temperature dependence in the experimental data (Fig. \ref{fig:ds}b) is much faster, but we are able to qualitatively reproduce the observation that the depletion is largest at the lowest energy. 

We now turn to the possible interpretation of our experimental results. The key observation is that the low temperature optical conductivity features a residual metallic contribution. This was observed previously in non-equilibrium experiments \cite{Stojchevska2014, Han2015, Gerasimenko2019, Ravnik2021, Gao2021, Venturini2022} and has been attributed to the formation of domain walls \cite{Ma2016,Cho2016,Gerasimenko_npj_2019} arising from a collapse of layer dimerization \cite{Stahl2020}. Furthermore, it has been shown that these domain walls can be 'charged' \cite{Salzmann2022}. We propose that domain walls or stacking faults in our crystals are pervasive throughout the bulk of the crystal and collectively contribute to the observed optical conductivity. As temperature increases, thermal fluctuations lead to a removal of these domain walls and the long range CCDW order with a full optical gap can set in. 

The analysis of our optical data shows that there is a distinct difference between the electronic mechanism driving the formation of the mosaic and long range ordered CCDW state. The former is accompanied by a depleted density of states around the Fermi level. Such a gradual depletion is often referred to as a 'pseudogap' in the context of the cuprate superconductors. Its origin in the cuprate case is unknown, but some form of pair formation without long range coherence \cite{eagles1969} has been suggested as a possible source. In TaS$_{2-x}$Se$_{x}$, this scenario may hold true in the mosaic phase where the onset of long range CCDW order is suppressed. The interplay between the metallic domain boundaries and short range CCDW order could be analogous to incoherent fluctuations of the pairing field, thus providing a route to the formation of 'preformed density fluctuations'. 

To summarize, we have observed a bulk, meta-stable phase in 1T-TaS$_{1.2}$Se$_{0.8}$ that closely resembles the non-equilibrium mosaic phase observed in 1T-TaS$_{2}$. This phase is accompanied by the formation of a large pseudogap but has a residual metallic component. By changing temperature, we can also reach the CCDW phase which instead has a full optical gap. It would be very interesting to further explore the connection to the non-equilibrium mosaic phase with diffraction, ARPES and STM experiments. 

\begin{acknowledgments}
EvH would like to especially thank Dr. Lev Gasparov for making spectral weight data of Ref. \cite{Gasparov2002} available for comparison with our work. EvH acknowledges support for this research from the research center for quantum software and technology, QuSoft. EDC acknowledges support from...
\end{acknowledgments}

\section{Supplementary Information}
Bulk 1T-TaS$_{2-x}$Se$_{x}$ crystals with a nominal stoichiometry of x\,=\,0.8 and x\,=\,1.0 were grown by the chemical vapor transport method . The starting materials were weighed in a nitrogen-filled glove-box using an Ohaus Pioneer digital balance with 0.5 mg precision. The elements were sealed in a quartz ampoule in vacuum. During the evacuation, it is necessary to place the bottom of the ampoule (containing the powdered elements) within a bucket of dry ice, to ensure that the iodine powder does not sublime and escape the ampoule before sealing. The sealed ampoule was placed in the middle of a Lenton two-zone furnace, which was then sealed at either end using insulating blocks. The furnace was ramped up from room temperature at 0.5 K/min to final temperatures of 1173 K and 1233 K for the growth and reaction zones, respectively. The growth proceeded for a total of seven days,
at the end of which time the ampoule was removed and quenched in water. The single crystals thus obtained range from one to a few mm in size and are stored in a nitrogen-filled glove-box to minimise oxidation.

\begin{figure*}[thb]
    \includegraphics[width=1.99\columnwidth]{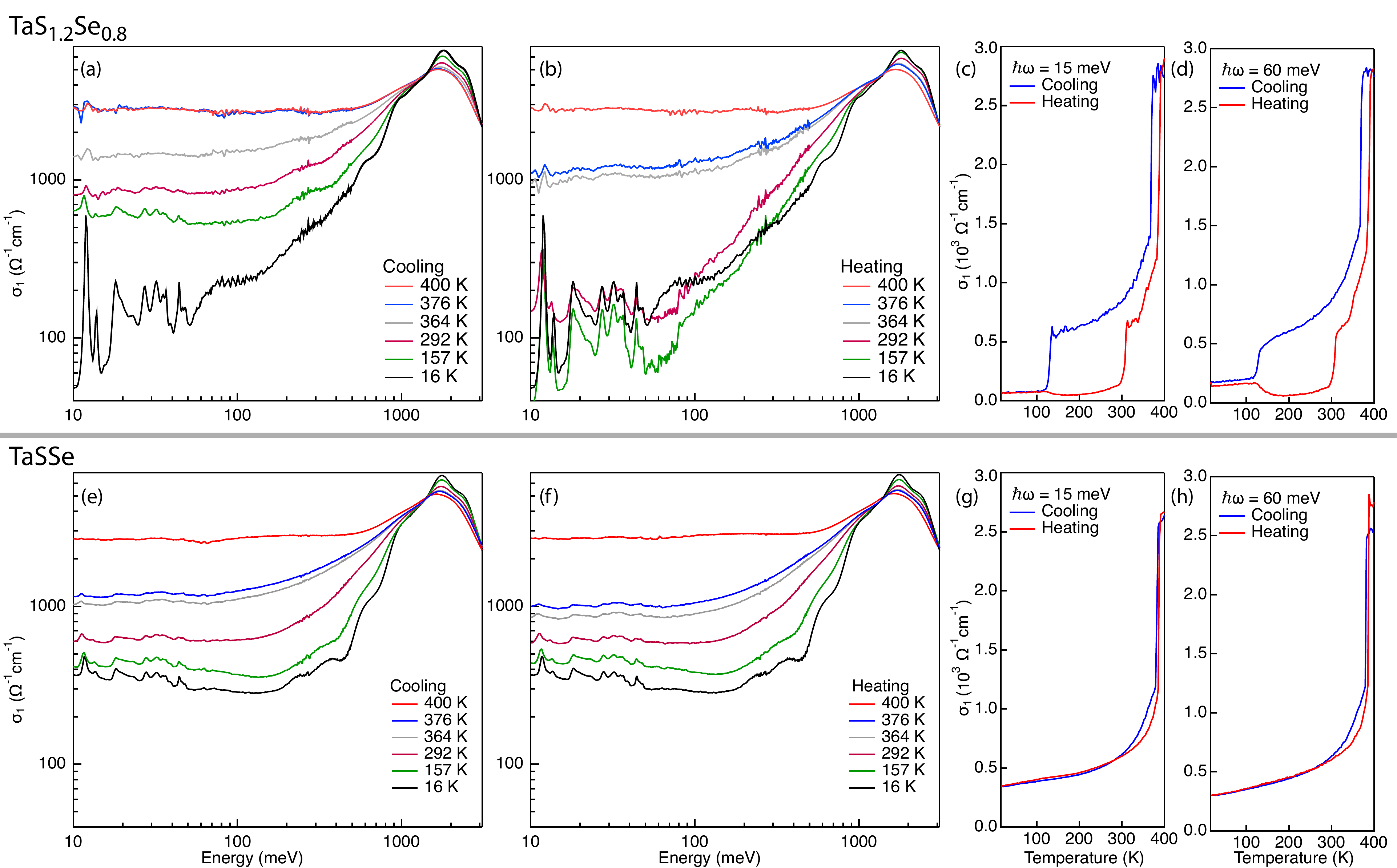}
    \caption{Optical conductivity of Se doped TaS$_{2-x}$Se$_{x}$. Top row shows the data for x\,=\,0.8 and bottom row for x\,=\,1.0. (a): $\sigma_{1}(\omega,T)$ for selected temperatures, measured during cooling. As temperature decreases a series of IR active phonon modes become visible. (b): $\sigma_{1}(\omega,T)$ for selected temperatures, measured during heating. The data shows clear spectral differences compared to the cooling data. (c,d): the temperature dependence at two selected frequencies clearly show distinct cooling and heating transitions that are not simply different due to hysteresis of the transitions. Notably, panel (d) shows two distinct transitions while cooling and 4 transitions while heating. Panels (e-g) show the same results as (a-d) except that in the x\,=\,1.0 sample the mosaic and CCDW phase are suppressed.}
    \label{fig:sups1}
\end{figure*}
In addition to the x\,=\,0.8 crystal we have also performed extensive experiments on a x\,=\,1.0 crystal. This crystal does not display signatures of the mosaic phase or CCDW phases and provides a good contrast with the behaviour of the x\,=\,0.8 crystal. Fig. \ref{fig:sups1} shows a comparison between 1T-TaS$_{1.2}$Se$_{0.8}$ (top panels) and 1T-TaSSe (bottom panels). The overal spectral features are very similar. We observe a large temperature dependent interband conductivity also in this crystal, with an interband response that is quantitatively the same. This provides some evidence that this large change in interband conductivity is not directly related to the CCDW or mosaic phase transitions. 

\bibliography{paper_lib.bib}
\end{document}